\documentclass{PoS}

\usepackage{amsfonts, amssymb, amsmath, graphicx, color}

\newcommand{\beq}{\begin{equation}}
\newcommand{\eeq}{\end{equation}}
\newcommand{\calH}{ {\cal H} }
\newcommand{\rmd}{ {\rm d} }
\newcommand{\rmi}{ {\rm i} }
\newcommand{\rme}{ {\rm e} }

\newcommand{\calG}{ \mathcal{G} }

\newcommand{\Tr}{{\rm Tr}}

\newcommand{\Nc}{N_{\rm c}}

\newcommand{\lqcd}{\Lambda_{\rm QCD}}
\newcommand{\vp}{\vec{p}}

\title{Phenomenological QCD equations of state for neutron star mergers}
\ShortTitle{Phenomenological QCD equations of state for neutron star mergers}

\author{\speaker{Toru Kojo}\\%
      Central China Normal University\\
      E-mail: \email{torujj@mail.ccnu.edu.cn}}
\abstract{We delineate the properties of dense QCD matter through equations of state constrained by the neutron star observations. The two solar mass constraint, the radius constraint of 11-13 km, and the causality constraint on the speed of sound, are used to develop the picture of hadron-quark continuity in which hadronic matter continuously transforms into quark matter. A unified equation of state at zero temperature and $\beta$-equilibrium is constructed by a phenomenological interpolation between nuclear and quark matter equations of state. For applications to supernovae and neutron star mergers, the unified equation of state is perturbed by temperature corrections. }
\FullConference{Critical Point and Onset of Deconfinement\\
		 7-11 August, 2017\\
		 The Wang Center, Stony Brook University, Stony Brook, NY}
\begin{document}	

\section{Introduction}

The study of the QCD phase structure at large baryon density has been a difficult problem, partly because the lattice Monte Carlo simulations based on the QCD action are not at work, and partly because many-body problems with strong interactions are very complex in theoretical treatments. Currently the best source of information for dense QCD is the physics of neutron stars from which one can extract useful insights into QCD equations of state and transports beyond the nuclear regime \cite{Baym:2017whm}. The physics of neutron stars include the structure of neutron stars of low temperature matter in $\beta$-equilibrium, and supernovae as well as neutron star mergers which contain hot QCD matter with various lepton fractions and neutrinos. The domain relevant for these physics is the baryon density of $n_B \sim 1-10n_0$ ($n_0 \simeq 0.16{\rm fm}^{-3}$: nuclear saturation density) or baryon chemical potential of $\mu_B\sim 1-2\, {\rm GeV}$, and the temperature of $T\sim 0-100\,{\rm MeV}$. 

In this talk I discussed on-going attempts to delineate the properties of dense QCD matter using the astrophysical constraints. There are remarkable progress in observations that constrain our understanding on the nature of matter. They include the discoveries of two-solar mass ($2M_\odot$) neutron stars \cite{2m_1}, the constraints for the neutron star radii from X-ray analyses \cite{Ozel:2016oaf,Steiner2016}, and most remarkably, the detection of the gravitational waves \cite{GW170817} and the electromagnetic signals \cite{GW170817A} from neutron star mergers found on August 17 just after this CPOD meeting.

The first part of this talk is devoted to the discussions of neutron star structure and its implications for the properties of QCD matter at zero temperature. In particular we set up a quark model description for the high density part. Then we apply the model to descriptions of matter at finite temperature, which is the second topic in this talk.

\section{QCD matter at zero temperature in $\beta$-equilibrium}

Neutron star observations give us useful constraints on equations of state through the neutron star mass-radius ($M$-$R$) relations. The mass and radius become larger for stiffer (larger pressure $P$ at given energy density $\varepsilon$) equations of state since the pressure pushes back matter attracted by the gravity. In principle, a precisely determined $M$-$R$ relation can be used to directly reconstruct the neutron star equations of state \cite{lindblom}. While the precision at present is not good enough for the direct inversion procedure to determine the equations of state, it has been known \cite{radius} that the shape of $M$-$R$ curves can be characterized by equations of state at three characteristic regions in $n_B$ (Fig.1) 
At low density, $n_B \lesssim 2n_0$, $R$ rapidly decreases as $M$ increases, and around $\sim 2n_0$, the $M$-$R$ starts to go vertically without much change in $R$. Then the curve reaches the maximum in $M$ at $n_B \gtrsim 5n_0$. Using these correlations between $M$-$R$ and $n_B$, one can focus on the radius constraint for the low density equations of state, or for the high density part one can focus on the maximum mass.

\begin{figure}
\begin{center}
\vspace{-1.0cm}
\hspace{-2.5cm}
\includegraphics[width=6.0cm,bb=0 0 600 600, angle =270]{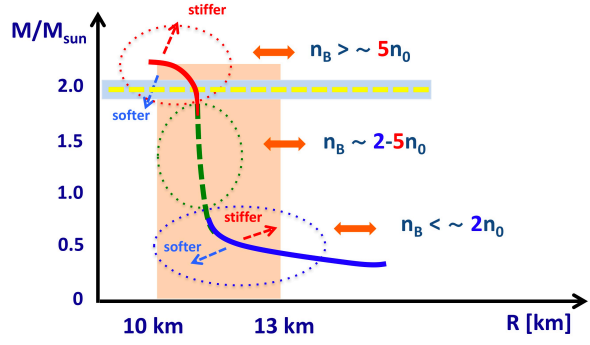}
\vspace{-0.3cm}
\caption{
\footnotesize{ The correlation between the $M$-$R$ relation and equations of state.
}
  }
  \vspace{-0.5cm}
  \end{center}
 \label{fig:M-R}
\end{figure}

One of the established constraint is the existence of two-solar mass ($2M_\odot$) neutron stars  \cite{2m_1}, which tells us that high density equations of state at $n_B\gtrsim 5n_0$ should be stiff enough to prevent stars at $M \simeq 2M_\odot$ from collapsing to a blackhole. Another important constraint comes from the estimate of $R$, most typically $R_{1.4}$ for $1.4M_\odot$ stars, and it tells us whether low density equations of state are stiff or not. There have been many predictions for $R_{1.4}$ which ranges from $\simeq 10$ km to $\simeq 16$ km. Below equations of state giving $R_{1.4} \lesssim 13$ km will be called soft low density equations of state, otherwise regarded as stiff one. The estimate of $R$, which has been based on spectroscopic analyses of the X-rays from the neutron star surface, includes more systematic uncertainties than in the mass determination. But the current trend on these analyses converges toward the estimate $R=11-13$ km, which is consistent with microscopic nuclear calculations at low density \cite{Akmal:1998cf,Togashi:2017mjp}. Moreover, as we will discuss later, the analyses of gravitational waves from neutron star mergers favor soft low density equations of state.

\begin{figure}
\begin{center}
\vspace{-1.0cm}
\hspace{-2.5cm}
\includegraphics[width=8.0cm,bb=0 0 600 600, angle =270]{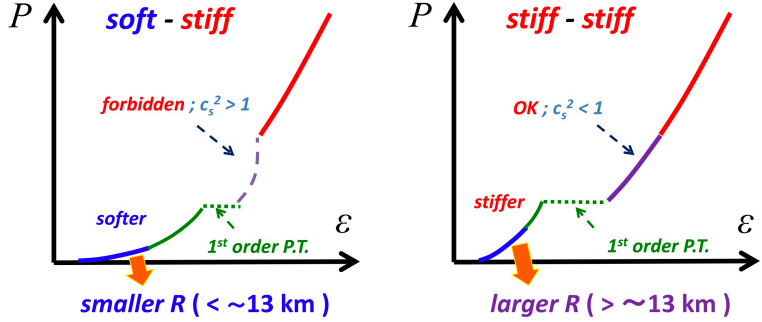}
\vspace{-1.2cm}
\caption{
\footnotesize{ The pressure  v.s. energy density for soft-stiff and stiff-stiff equations of state. The slope is given by $\partial P/\partial \varepsilon = c_s^2$, the sound speed square, which must be smaller than $1$. The soft-stiff combination of low and high density equations of state disfavors the strong 1st order phase transition and has the radius smaller than the stiff-stiff combination.
}
  }
  \vspace{-0.0cm}
  \end{center}
 \label{fig:1st}
\end{figure}

These somewhat independently constrained high and low density equations of state must be put together, and at this stage the causality constraint plays a very important role, especially when we try to construct soft-stiff equations of state in which $P(\varepsilon)$ is small at low density but large at high density. In order to connect the soft equations of state to stiff ones, there must be a domain such that $\partial P/\partial \varepsilon = c_s^2$, the sound speed square, is large, but it should not exceed the light velocity. This constrains the structure of $P(\varepsilon)$ for $2n_0 \lesssim n_B \lesssim 5n_0$, and the strength of possible 1st order phase transitions in particular. Indeed if we try to put the 1st order phase transition by hand in the intermediate region  (Fig.2), 
there is a jump in $\varepsilon$ for fixed $P$, therefore after the phase transitions we need even larger $\partial P/\partial \varepsilon$ to get the connection to the high density part in $P(\varepsilon)$. For this reason below we will focus on the equations of state which do not contain any strong 1st order phase transitions from $2n_0$ to $5n_0$, although small 1st order phase transitions are still not excluded in principle. More systematic analyses are given in Ref.\cite{Alford:2013aca}.

Here we note that at $n_B \gtrsim 5n_0$ the baryons start to touch each other provided that the core radius is $\sim 0.5\,{\rm fm}$. The Fermi momentum of quarks in 3-flavor matter at $n_B=5n_0$ corresponds to $p_F \simeq 400\,{\rm MeV}$ which is larger than the QCD nonperturbative scale $\Lambda_{{\rm QCD}} \simeq 200\,{\rm MeV}$. Meanwhile, purely hadronic calculations are not under theoretical control beyond $n_B \gtrsim 2n_0$, due to large corrections from higher order effects such as rapidly growing many-body forces in nuclear calculations, the appearance of new degrees of freedom other than nucleons, etc. Thus we expect the importance of quark substructure effects in neutron star matter for $n_B \gtrsim 2n_0$, and further imagine the formation of quark matter around $n_B \simeq 5n_0$.  Combining this microscopic insights with the assumptions on the absence of strong 1st order phase transitions, we view the QCD matter from the picture of hadron-quark continuity \cite{Schafer:1998ef,Hatsuda:2006ps}, in which hadronic matter smoothly transforms into quark matter.

\begin{figure}
\begin{center}
\vspace{-1.0cm}
\hspace{-.cm}
\includegraphics[width=3.5cm, bb=0 0 300 800, angle =270 ]{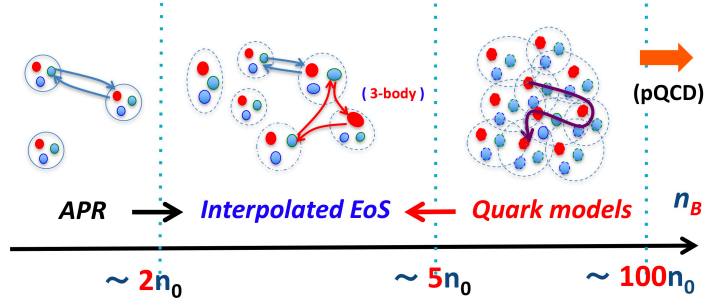}
\vspace{2.2cm}
\caption{
\footnotesize{ The 3-window modeling of the QCD matter. 
}
  }
  \vspace{-0.2cm}
  \end{center}
 \label{fig:3window}
\end{figure}

Based on the picture of the hadron-quark continuity, we construct QCD equations of state by a 3-window approach (Fig.3) 
\cite{Masuda2013-0,Kojo2014}. At low density, $n_B \lesssim 2n_0$, the matter is dilute and baryons are well-defined objects, so the equations of state are described by nuclear ones. We use here the Akmar-Phandheripande-Ravenhall (APR) equation of state as a representative \cite{Akmal:1998cf}. At high density, $n_B\gtrsim 5n_0$, the matter is dense enough for a quark Fermi sea to form, so the equations of state are described by quark matter ones. This domain is parameterized by microscopic interaction parameters in a schematic Nambu-Jona-Lasinio (NJL) type quark model for hadron physics. In between there is a matter for which neither purely hadronic nor quark matter descriptions are appropriate, so, applying the hadron-quark continuity picture, we interpolate the APR and quark model equations of state. Specifically our interpolation is done with polynomials
\beq
P(\mu_B) = \sum_{n=0}^5 c_n \mu_B^n \,.
\eeq
To determine the coefficients $c_n$'s, we first compute $n_B=\partial P/\partial \mu_B$, and then demand, at $n_B=2n_0$ and $5n_0$, the interpolating function to match with the APR and quark equations of state up to the second derivative.

In this phenomenological modeling, we need to choose a quark model for $n_B \gtrsim 5n_0$. Guided by the continuity picture, the form of effective models is inspired from those for hadron physics. Here relatively short range interactions are important as the quark matter regime observes the contents inside of hadrons. Those include the physics of chiral symmetry breaking and color-magnetic interactions whose relevant scales, $0.2-1\,{\rm GeV}$, are a bit larger than the scale of confinement, $\lqcd \sim 0.2\, {\rm GeV} \sim 1{\rm fm}^{-1}$. Our effective Hamiltonian is ($\mu_q =\mu_B/3$)
\begin{align}
\calH  
= \bar{q} (\rmi \gamma_0 \vec{\gamma}\cdot \vec{\partial} + m -\mu_q \gamma_0)q 
- \frac{G_s}{2} \sum^8_{i=0} \left[ (\overline{q} \tau_i q)^2 + (\bar{q} \rmi \gamma_5 \tau_i q)^2 \right] 
+ 8 K ( \det\,\!\!_{\rm f} \bar{q}_R q_L + \mbox{h.c.}) \nonumber \\
+ \calH_{ {\rm conf} }^{ {\rm 3q\rightarrow B} }  - \frac{H}{2} \!\sum_{A,A^\prime = 2,5,7} \!
 \left(\bar{q} \rmi \gamma_5 \tau_A \lambda_{A^\prime} C \bar{q}^T \right) \left(q^T C \rmi \gamma_5 \tau_A \lambda_{A^\prime} q \right) + \frac{G_V}{2} (\overline{q} \gamma^\mu q)^2
 \,.
\end{align}
The first line is the standard NJL model with $u,d,s$- quarks and responsible for the chiral symmetry breaking. We use the Hatsuda-Kunihiro parameter set \cite{Hatsuda1994} with which the constitutent quark masses are $M_{u,d} \simeq 336\,{\rm MeV}$ and $M_s \simeq 528\,{\rm MeV}$. The second line includes the confining interactions which trap 3-quarks into a baryon, the color magnetic interaction for color-flavor-antisymmetric S-wave interaction which is attractive, and phenomenological vector repulsive interactions which are inspired from the $\omega$-meson exchange in nuclear physics. Actually we will not explicitly treat the confining term; instead we restrict the use of this model to the high density region where baryons overlap.

We note that while the form of the Hamiltonian is obtained by extrapolating the description of hadron and nuclear physics, in principle the range of parameters $(G_s, K, g_V, H)$ at $n_B \gtrsim 5n_0$ can be considerably different from those used in hadron physics. We use the neutron star constraints to examine the range of these parameters and from which we delineate the properties of QCD matter at $n_B \gtrsim 5n_0$. Below we vary ($g_V, H)$, while assume that $(G_s, K)$ do not change from the vacuum values appreciably; this assumption will be checked posteriori. The medium modifications of bare coupling was demonstrated in Ref.\cite{Fukushima:2015bda}.

Our Hamiltonian for quarks, together with the contributions from leptons, is solved within the mean field (MF) approximation. The neutrality conditions for electric and color charges, as well as the $\beta$-equilibrium condition, are imposed.
In the MF treatments we find that the chiral and diquark condensates coexist at $n_B \gtrsim 5n_0$. For the range of parameters we have explored, the diquark pairing always appears to be the color-flavor-locked (CFL) type at $n_B \gtrsim 5n_0$; other less symmetric pairings such as the 2SC type appear only at the lower density where the confining effects are not negligible.

Now we examine the roles of effective interactions by subsequently adding $g_V$ and then $H$ to the standard NJL model. First of all, in order to make equations of state stiff, $(G_s, K)_{@5n_0}$ should remain comparable to the size of its vacuum values; the large reduction of these parameters accelerates the chiral restoration that yields contributions similar to the bag constant, i.e., the positive (negative) contributions to energy (pressure). As a result the significant softening takes place in equations of state. Actually even if we fix $(G_s, K)_{@5n_0}$ to the vacuum values, the strong 1st order chiral transition takes place at $n_B \sim 2-3n_0$ in the standard NJL model, so the equations of state at $n_B\gtrsim 5n_0$ is too soft to pass the $2M_\odot$ constraint. 

This situation is changed by adding $g_V$. It stiffens the equations of state in two-fold ways. Firstly the repulsive interactions obviously contribute to the stiffening. Secondly, it delays the chiral restoration by tempering the growth of baryon density as a function of $\mu_B$, so that there are no radical softening associated with the chiral restoration. In fact the 1st order transition turns into a crossover in the range of $g_V$ we explored. The value of $g_V$ large enough to pass the $2M_\odot$ constraint, however, causes another kind of problem in connecting the APR and quark model pressure, see the left panel of Fig. 4. ; with larger $g_V$ quark pressure $P(\mu_B)$ tends to appear at higher $\mu_B$ with the less slope, and as a consequence the pressure curve in the interpolation region tends to contain an inflection point at which $\partial^2 P/(\partial \mu_B)^2$ is negative. Such region is thermodynamically unstable so must be excluded. Therefore while larger value of $g_V$ is favored to pass the $2M_\odot$ constraint, it generates more mismatch between the APR and quark pressure in the $\mu_B$ direction.

\begin{figure}
\begin{center}
\vspace{-2.5cm}
\hspace{-.cm}
\includegraphics[width=5.0cm, bb=0 0 300 800, angle =270 ]{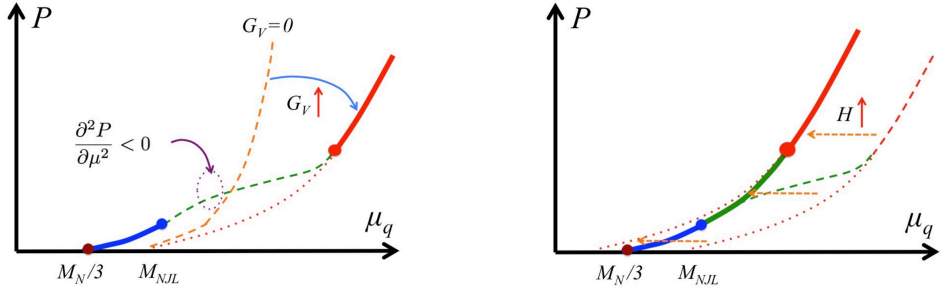}
\vspace{2.2cm}
\caption{
\footnotesize{ The impacts of the vector and color-magnetic interactions.
}
  }
  \vspace{-0.2cm}
  \end{center}
 \label{fig:vector-diquark}
\end{figure}

Here the color magnetic interactions improve the situation, see the right panel of Fig. 4. We note that the onset chemical potential of the APR pressure is the nucleon mass $\mu_B \simeq 939\,{\rm MeV}$, while for the NJL pressure it is $\mu_B\simeq 3M_{u,d} \simeq 1018\, {\rm MeV}$. In conventional picture of quark models, the nucleon and $\Delta$ masses are split by the color-magnetic interaction, and the nucleon mass is reduced from $3M_{u,d}$. From this viewpoint, the color magnetic interactions induce the 
overall shift of the NJL pressure toward the lower chemical potential, thus make the matching between the APR and quark pressure curves much better.

The $M$-$R$ relations are shown in Fig.5 
for the parameter sets $(g_V, H)/G_s = (0.5,1.4), (0.8,1.5)$, and $(1.0, 1.6)$. For all these sets, the radius of a neutron star at the canonical mass $1.4M_\odot$ is $11.3-11.5$km, mainly determined by our APR equations of state. In these sets, only the set $(0.8,1.5)$ fulfils the all constraints; the set $(0.5,1.4)$ is slightly below the $2M_\odot$ constraint, while $(1.0, 1.6)$ slightly violates the causality bound. More exhaustive parameter surveys \cite{Baym:2017whm} show that $g_V$ should be $ \gtrsim 0.7 G_s$, and $H \gtrsim 1.4 G_s$ which are comparable to the vacuum scalar coupling. For given $g_V$ the value of $H$ is fixed to $\sim10$\%; in fact we do not have much liberty in our choice when we connect the APR and quark matter pressures.

 With such strong effective couplings, we expect that gluons in the non-perturbative regime still survive in spite of the presence of quark matter, as discussed in Ref.\cite{McLerran:2007qj} by using the picture of $1/\Nc$ expansion. In two space-time dimensions such a state of matter is indeed possible \cite{Kojo:2011fh}.

\begin{figure}
\begin{center}
\vspace{-1.0cm}
\hspace{-2.5cm}
\includegraphics[width=8.0cm,bb=0 0 600 600, angle =270]{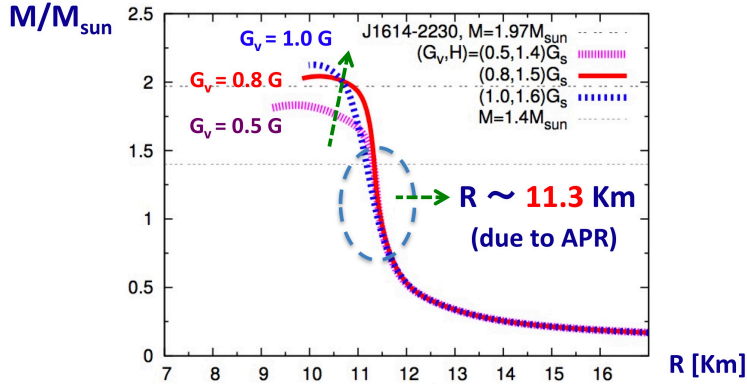}
\vspace{-1.2cm}
\caption{
\footnotesize{ The mass-radius relations from the 3-window equations of state for sets of parameters, $(g_V,H)/G_s=(0.5,1.4), (0.8,1.5), (1.0,1.6)$. Only the set $(0.8,1.5)$ satisfies the $2M_\odot$ and causality constraints.
}
  }
  \vspace{-0.0cm}
  \end{center}
 \label{M-R_HQC}
\end{figure}

\section{Toward equations of state for neutron star mergers}

The zero temperature equations of state constructed in our framework is a result of fits to neutron star constraints, with $(g_V, H)$ as parameters. To check the validity of the descriptions, it is desirable to calculate quantities which sensitively depend on the microscopic picture. Below we consider the quantities sensitive to the excitations as good measures for the phase structure and the symmetry breaking patterns. 

The thermal equations of state are such example, and have impacts on the dynamics of supernovae explosions and neutron star merger events. Of particular concern in our framework is the quark matter part which has been most uncertain. The supernovae matter \cite{Janka:2012wk} is probably not dense enough to study the impact of quark matter, since the maximum density is likely to be $n_B \simeq 2-3n_0$ close to the nuclear regime. In neutron star mergers \cite{baiotti} we have more chances; the maximum baryon density can be as high as $\sim 5n_0$. The temperature distributions calculated in dynamical simulations with hadronic equations of state suggest that the hottest domain of matter is $\sim 2n_0$, having the temperature of $\sim$20-100 MeV, while the densest part has the lower temperature of $\sim$10-20 MeV. We note that all these estimates depend on equations of state and therefore may vary if we consider equations of state with very different thermal properties.

\begin{figure}
\begin{center}
\vspace{-2.cm}
\hspace{-2.6cm}
\includegraphics[width=9.0cm,bb=0 0 600 600, angle =270]{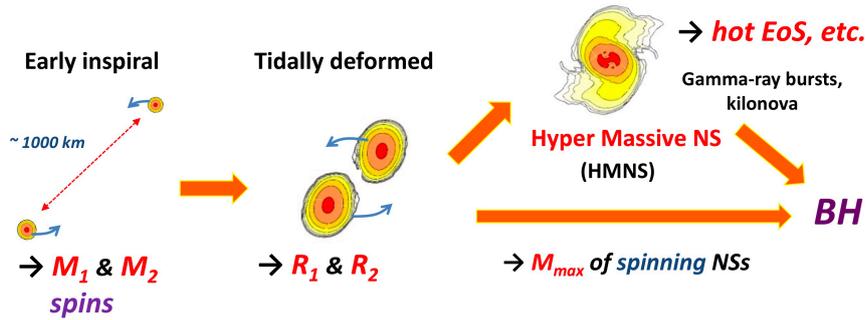}
\vspace{-2.0cm}
\caption{
\footnotesize{ 
The time evolution of a neutron star binary.
}
  }
  \vspace{-0.2cm}
  \end{center}
 \label{fig:NSbinary}
\end{figure}

Here we briefly outline the time evolution of neutron star mergers (Fig.6). 
Each stage offers us different information through the gravitational wave (GW) emissions and electromagnetic (EM) signals. Neutron star binary systems can be regarded as the time varying quadratic poles that are the sources of gravitational waves. When two neutron stars are largely separated, the two neutron stars can be treated as point particles and the amplitudes of GWs are small. This stage is called inspiral phase. GWs from the inspiral can be precisely calculated by the post-Newtonian approximation, in which $v/c$ ($v$: the velocity of relative motion) is treated as a small expansion parameter. In principle one can study not only the masses of two neutron stars but also the spins since the spin-orbit and spin-spin interactions appear in higher orders of the expansion \cite{Cutler:1994ys}. As two neutron stars approach, the inspiral phase changes into the tidal deformed phase, where the internal structure of each neutron star starts to be relevant in the waveform of GWs \cite{Hinderer:2009ca}. The particularly important characteristic quantity is the tidal deformability of a star whose value is strongly correlated with the compactness $M/R$ of the star. By measuring the mass of the star from the inspiral phase, one can then constrain the size of the radius. Therefore the inspiral and tidal deformed phases tell us a lot about the structure of neutron stars at zero temperature before the coalescence. 

Eventually two neutron stars coalesce. The produced object either promptly collapses into a blackhole, or remains for $\sim 10$ms as a hypermassive neutron star with large differential rotation and the thermal pressure. The physics in the coalescence regime is a highly dynamical problem which requires sophisticated numerical simulations \cite{baiotti} on general relativistic effects and transports for given equations of state. The major uncertainty arises from the QCD equations of state at $n_B\gtrsim 2n_0$ at finite temperature and various lepton fractions. The temperature of the matter is raised by absorbing the heat generated through the friction of two neutron stars.

As for finite temperature equations of state, the standards have been the Lattimer-Swetsy (LS) \cite{Lattimer:1991nc} and Shen equations of state \cite{Shen:1998gq}. Nowadays more equations of state are constructed, reflecting the recent progress in astrophysical constraints \cite{Togashi:2017mjp,Steiner:2012rk}. On the other hand, almost all of them are based on the hadronic descriptions, so the extrapolation of those equations of state beyond $\sim 2-3n_0$ may be questionable. There are few finite temperature quark equations of state, and they are based on a bag model, or perturbative QCD calculations \cite{Kurkela:2016was}, or a 3-window modeling with a quark model \cite{Masuda:2015wva}. In these treatments the most relevant excitations are gapless quarks whose contributions to equations of state are much larger than those from gluons, because the larger phase space, $\propto 4\pi p_F^2$, is available for quarks near the Fermi sphere. A quark model study based on the 3-window approach suggests that for a supernova matter in the isentropic condition, the temperature of the matter in quark matter description can be 3-4 times smaller than purely hadronic descriptions since quarks with large phase space can carry large entropy even at low temperature.

Considering the pairing effects, however, one can imagine alternative scenarios in which thermal properties are qualitatively different from the gapless quark matter. In the CFL phase quarks and gluons are both massive so that thermal contributions appear much more suppressed than in the gapless quark matter \cite{Alford:2007xm}. For our parameter sets used for the zero temperature equations of state, we have the CFL phase at $n_B\gtrsim 5n_0$ and the size of diquark gaps are $\sim 200$ MeV. So almost no quarks absorb heat. Other possible candidates of thermal excitations are lepton pairs, neutrinos, and collective modes. The leptons do not enjoy the Fermi surface enhancement because the CFL has equal number of $u,d,s$ quarks and no electrons are necessary for charge neutrality. So they appear only as lepton-antileption pairs.

\begin{figure}
\begin{center}
\vspace{-3.5cm}
\hspace{-2.9cm}
\includegraphics[width=11.0cm,bb=0 0 600 600, angle =270]{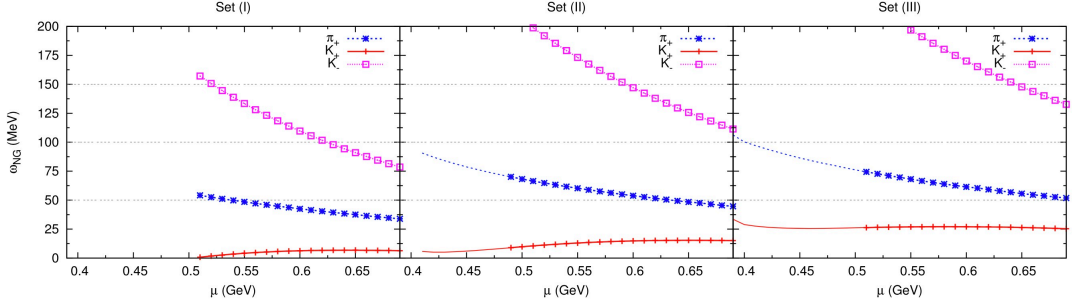}
\vspace{-3.0cm}
\caption{
\footnotesize{ 
The excitation energies of NG modes at rest. The bold line part is given for $n_B \gtrsim 5n_0$. 
}
  }
  \vspace{-0.2cm}
  \end{center}
 \label{fig:NGmodes}
\end{figure}

As for the contributions of collective modes, the leading contributions should be the Nambu-Goldstone (NG) bosons \cite{Son:1999cm}. If the current masses and electric charges for (u,d,s)-quarks were zero, there would be $SU(3)_L \times SU(3)_R \times SU(3)_c \times U(1)_B \times U(1)_{em} $ symmetry, broken into $SU(3)_{c+L+R} \times U(1)_{Q'}$ in the CFL phase. The resulting NG modes are 9 massless modes and 8 gluons are massive. If  $U(1)_A$ explicit breaking is suppressed by medium effects, then one can view $U(1)_A$ as if an ordinary symmetry plus the small violation for which the picture of one more pseudo-NG mode is at work. In neutron stars, however, the explicit breaking associated with the current quark masses and charges are important; only one NG mode associated with $U(1)_B$ is massless while the others are all pseudo-NG modes. Shown in Fig.7 
are the spectra of $\pi^+,K^+$, and $K^-$ for $(g_V,H)/G_s=(0.5,1.4), (0.8,1.5), (1.0,1.6)$ for the set (I), (II), and (III). They were computed within the random phase approximation (RPA) on top of the MF background used for the zero temperature neutron star equations of state \cite{Kojo:2016dhh}. We found that the isospin remains good symmetry so here do not show $\pi^-, K_0,\bar{K}_0$. We also computed the 3-neutral NG modes which are linear combinations of $\pi_0,\eta,\eta'$ quantum numbers. 

The overall tendency is that (i) as we increases the coupling, the resulting spectra of NG modes appear at higher energies; (ii) the typical excitation energies are $\sim 50-200$ MeV, except $K_+$ and $K_0$ modes; (iii) $K_+$ and $K_0$ are anomalously light because of the effective chemical potential associated with the imbalance for $u,d$ and $s$ -quarks; (iv) at weaker coupling there may be a kaon-condensation at $n_B\gtrsim 5n_0$, while in our setup for neutron stars it did not occur. Earlier calculations based on the NJL model can be found in Ref.\cite{Kleinhaus:2007ve}, while our study updated the results taking the recent constraints into account.

From these analyses, we expect that thermal contributions from NG modes are like those in the zero density case where the lightest hadrons are pions with the masses $\simeq 140$ MeV. On the other hand, recent studies on the hadron resonance gas around $T\simeq 150$ MeV suggest that, even when the massive thermal excitations are suppressed by the Boltzmann factors, the sum of those contributions may give significant contributions \cite{Ding:2015ona}. So it is desirable also at finite density to include not only the lightest modes but also excitations at higher energies into thermodynamic potentials.

For this reason we evaluate the thermal contributions using the phase shift representation \cite{Beth:1937zz,Dashen:1969ep,Zhuang:1994dw}, which summarizes the correlations from low to high energies. Specifically we consider 2-body channels, including the correlated pairs made of particles, holes, antiparticles. They are computed at the level of the RPA. For a channel with a quantum number $X$,
\beq
\Omega^{ {\rm 2-body} }_X =  \int \! \frac{\, \rmd\vp\,}{\, (2\pi)^3 \,} 
\int_{-\infty}^\infty \frac{\, \rmd \omega \,}{2\pi} \left[ \frac{\, |\omega| \,}{2} + T \ln \left(1-\rme^{ -|\omega|/T } \right) \right] \frac{\, \rmd \delta_X(\omega,\vp) \,}{\rmd \omega} \,,
\label{eq:thermo_2body}
\eeq
where the all information in the channel $X$ (including chemical potentials) is encoded in the phase shift $\varphi_X$, which appears in the ratio of full $(\calG)$ and disconnected ($\calG_0$) 2-particle Green's functions,
\beq
\calG/\calG_0 = | \calG /\calG_0 | \rme^{\rmi \delta_X} \,.
\eeq
The phase shift representation of the thermodynamic potential looks the same as the hadron resonance gas if we substitute the phase shift for a bound state, $\delta_X (\omega,\vp) = \pi \theta \left( \omega -E_X (\vp) \right) $, in which the phase shift jumps by $\pi$ at the bound state energy.

\begin{figure}
\begin{center}
\vspace{-2.5cm}
\hspace{-2.5cm}
\includegraphics[width=8.0cm,bb=0 0 600 600, angle =270]{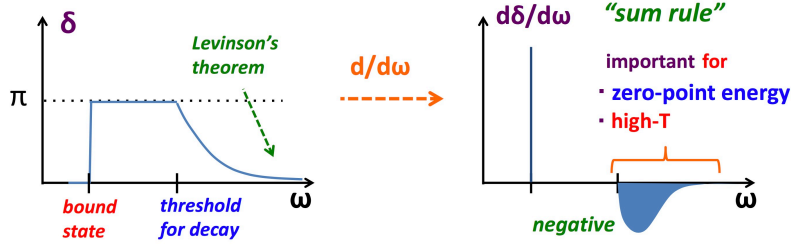}
\vspace{-1.8cm}
\caption{
\footnotesize{ An example of the phase shift and its derivative as a function of $\omega$.
}
  }
  \vspace{-0.3cm}
  \end{center}
 \label{phase_shift}
\end{figure}

As we noted, the phase shift representation includes the resonating continuum, but a part of continuum contributions are already taken into account when we calculate the single particle contributions. So we have to be sure that the double counting is correctly avoided. The double counting is absent when the phase shift satisfies the Levinson's theorem,
\beq
\Tr \delta_X(\omega=\infty) - \Tr \delta_X(\omega=0) = 0 \,,
\eeq
where the trace runs over all possible 2-body states. This constraint follows from the conservation of the total number of states with and without interactions,
\beq
0 = \int_0^\infty \rmd \omega \Tr\left[ {\rm Im} \calG - {\rm Im} \calG_0 \right] 
= \int_0^\infty \rmd \omega \partial_\omega \Tr\left[ {\rm Im} \ln \calG^{-1}/\calG_0^{-1}  \right] 
\,,~~~~~~~\calG = \frac{1}{E-H}\,,
\eeq
where ${\rm Im} \calG$ counts the number of states, and the integration of them remains invariant after interactions are added. The constraint from the Levinson's theorem requires that the phase shift at $\omega  \rightarrow \infty$ must return to the original value at $\omega=0$. This means that at the bound state energies the phase shifts jump by $\pi$ with the positive values for the derivative of $\delta_X$, while at higher energies there must be a region where the derivative of $\delta_X$ takes the negative value, hence giving the negative contribution to Eq.(\ref{eq:thermo_2body}), see Fig.8. 
These contributions are particularly important when we consider the physics around the transition temperature; they tend to cancel the resonance gas contributions at low energies in the two-body part of the thermodynamics potential, and as a result single particle contributions such as those from quarks and gluons saturate the thermodynamics, as one intuitively expects.

To complete the above mentioned descriptions for the thermodynamics, we need to calculate the distributions of the phase shift for $\omega$-$|\vp|$ plane. This is an on-going subject. For now we are working on the 3-flavor limit in which quark bases can be analytically constructed and the calculations of the RPA are much simpler than the realistic setup. Some results on the distributions are shown in Fig.9.

\begin{figure}
\begin{center}
\vspace{-1.cm}
\hspace{-1.cm}
\includegraphics[width=8.0cm,bb=0 0 600 600, angle =270]{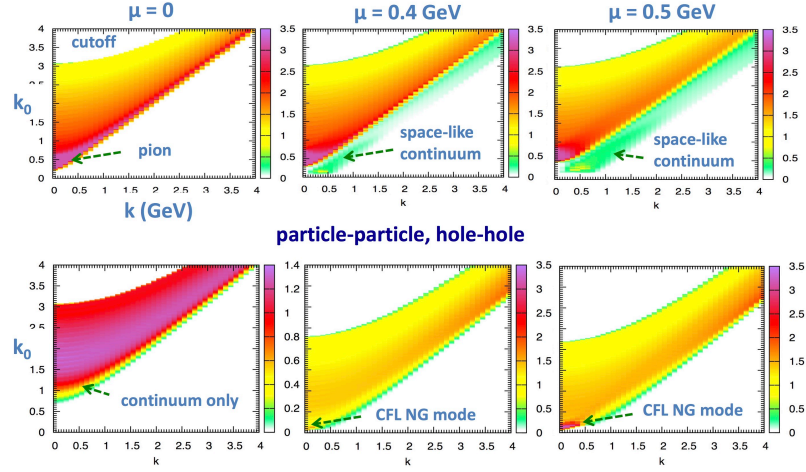}
\vspace{-0.5cm}
\caption{
\footnotesize{ An example of the phase shift distributions for pseudo-scalar channel in the $\omega-|\vp|$ plane, at quark chemical potential $\mu=0, 0.4,0.5$ GeV. The first line gives the particle-antiparticle result and the second for the particle-hole contributions. 
}
  }
  \vspace{-0.0cm}
  \end{center}
 \label{phase_shift}
\end{figure}

\section{Summary}

The physics of neutron stars are now giving significant constraints on the QCD equations of state. More observational constraints will come in next 10 years through the timing analyses of X-rays in the NICER program \cite{nicer}  and the GW detection by currently operating aLIGO, Virgo, GEO \cite{GEO}, and also KAGRA \cite{Kagra} under construction which will be ready soon. The electromagnetic counterparts associated with the GWs give the information about the ejecta, from which one can learn the dynamics at the coalescence regime. It is desirable to utilize all these information to improve our understanding on dense QCD matter.

The author thanks G. Baym, K. Fukushima, T. Hatsuda, P. Powell, Y. Song, T. Takatsuka for collaboration, and D. Blaschke for discussions about the phase shift representation. This work is supported by NSFC grant 11650110435.



\begin{thebibliography}{99}

\bibitem{Baym:2017whm}
A review,  G.~Baym, T.~Hatsuda, T.~Kojo, P.~D.~Powell, Y.~Song and T.~Takatsuka,
  arXiv:1707.04966 [astro-ph.HE].

%
\bibitem{2m_1}  
 Demorest, Paul et al., Nature {\bf 467} (2010) 1081-1083 arXiv:1010.5788 [astro-ph.HE];
J. Antoniadis et al., 2013, Science, {\bf 340}, 1233232.


\bibitem{Ozel:2016oaf}
  F.~Ozel and P.~Freire,
  Ann.\ Rev.\ Astron.\ Astrophys.\  {\bf 54} (2016) 401
  [arXiv:1603.02698 [astro-ph.HE]].

\bibitem{Steiner2016} A. W. Steiner, J. M. Lattimer, and E. F. Brown, 
Euro. Phys.. J, {\bf A52},  18:1-16  (2016).


\bibitem{GW170817} B. P. Abbott et al.  (LIGO Scientific Collaboration and Virgo Collaboration), 
Phys. Rev. Lett. {\bf 119}, 161101:1-18 (2017).

\bibitem{lindblom} L. Lindblom, ``Determining the nuclear equation of state from neutron-star masses and radii,"
Ap. J. {\bf 398},  569-573 (1992).

\bibitem{GW170817A} B. P. Abbott et al., 
Ap. J. Letters {\bf 848}, L12:1-59 (2017), and following papers in Ap. J. Letters {\bf 848}.

\bibitem{radius} J. M. Lattimer and M. Prakash, 
Phys. Rep. \textbf{442}, 109-65 (2007).


\bibitem{Alford:2013aca}
  M.~G.~Alford, S.~Han and M.~Prakash,
  Phys.\ Rev.\ D {\bf 88} (2013) no.8,  083013
  [arXiv:1302.4732 [astro-ph.SR]];


\bibitem{Schafer:1998ef}
  T.~Sch${\rm \ddot{a} }$fer and F.~Wilczek,
  Phys.\ Rev.\ Lett.\  {\bf 82} (1999) 3956
  [hep-ph/9811473].

\bibitem{Hatsuda:2006ps}
  T.~Hatsuda, M.~Tachibana, N.~Yamamoto and G.~Baym,
  Phys.\ Rev.\ Lett.\  {\bf 97} (2006) 122001
  [hep-ph/0605018];
  Z.~Zhang, K.~Fukushima and T.~Kunihiro,
  Phys.\ Rev.\ D {\bf 79} (2009) 014004
  [arXiv:0808.0927 [hep-ph]].






\bibitem{Akmal:1998cf}
  A.~Akmal, V.~R.~Pandharipande and D.~G.~Ravenhall,
  Phys.\ Rev.\ C {\bf 58} (1998) 1804
  [nucl-th/9804027].
  
  \bibitem{Togashi:2017mjp}
  H.~Togashi, K.~Nakazato, Y.~Takehara, S.~Yamamuro, H.~Suzuki and M.~Takano, 
  Nucl.\ Phys.\ A {\bf 961} (2017) 78.

\bibitem{Masuda2013-0} K. Masuda, T. Hatsuda, and T. Takatsuka, 
Astrophys. J. \textbf{764},12: 1-5 (2013); {\it ibid.}
Prog. Theor. and Exp. Phys. \textbf{7}, 073D01 (2013); {\it ibid.}
  Eur.\ Phys.\ J.\ A {\bf 52}, 65-79 (2016). 


\bibitem{Kojo2014}T. Kojo, P. D. Powell, Y. Song, and G. Baym, 
Phys. Rev. D \textbf{91}, 045003 (2015);
T.~Kojo, 
Eur.\ Phys.\ J.\ A {\bf 52},  51-69 (2016). 


\bibitem{Hatsuda1994}  T. Hatsuda and T. Kunihiro, 
Phys. Rep. \textbf{247}, 221-367 (1994).   

\bibitem{Fukushima:2015bda}
  K.~Fukushima and T.~Kojo,
  Astrophys.\ J.\  {\bf 817} (2016) no.2,  180
  [arXiv:1509.00356 [nucl-th]].

\bibitem{McLerran:2007qj}
  L.~McLerran and R.~D.~Pisarski,
  Nucl.\ Phys.\ A {\bf 796} (2007) 83
  [arXiv:0706.2191 [hep-ph]].

\bibitem{Kojo:2011fh}
  T.~Kojo,
  Nucl.\ Phys.\ A {\bf 877} (2012) 70
  [arXiv:1106.2187 [hep-ph]].


\bibitem{Janka:2012wk}
  A review, H.~T.~Janka,
  Ann.\ Rev.\ Nucl.\ Part.\ Sci.\  {\bf 62} (2012) 407
  [arXiv:1206.2503 [astro-ph.SR]].
  
  \bibitem{baiotti} A review, L. Baiotti and L. Rezzolla, 
  Repts. Prog. Phys. (in press) (2017), https://doi.org/
arXiv:1607.03540 [gr-qc].

  
\bibitem{Cutler:1994ys}
  C.~Cutler and E.~E.~Flanagan,
  Phys.\ Rev.\ D {\bf 49} (1994) 2658
  [gr-qc/9402014].
   


\bibitem{Hinderer:2009ca}
  T.~Hinderer, B.~D.~Lackey, R.~N.~Lang and J.~S.~Read,
  Phys.\ Rev.\ D {\bf 81} (2010) 123016
  [arXiv:0911.3535 [astro-ph.HE]].

\bibitem{Lattimer:1991nc}
  J.~M.~Lattimer and F.~D.~Swesty,
  Nucl.\ Phys.\ A {\bf 535} (1991) 331.
  
\bibitem{Shen:1998gq}
  H.~Shen, H.~Toki, K.~Oyamatsu and K.~Sumiyoshi,
  Nucl.\ Phys.\ A {\bf 637} (1998) 435
  [nucl-th/9805035].
  
\bibitem{Steiner:2012rk}
  A.~W.~Steiner, M.~Hempel and T.~Fischer,
  Astrophys.\ J.\  {\bf 774} (2013) 17
  [arXiv:1207.2184 [astro-ph.SR]].
  
\bibitem{Kurkela:2016was}
  A.~Kurkela and A.~Vuorinen,
  Phys.\ Rev.\ Lett.\  {\bf 117} (2016) no.4,  042501
  [arXiv:1603.00750 [hep-ph]].
\bibitem{Masuda:2015wva}
  K.~Masuda, T.~Hatsuda and T.~Takatsuka,
  PTEP {\bf 2016} (2016) no.2,  021D01
  [arXiv:1506.00984 [nucl-th]].
  
\bibitem{Alford:2007xm}
  M.~G.~Alford, A.~Schmitt, K.~Rajagopal and T.~Schäfer,
  Rev.\ Mod.\ Phys.\  {\bf 80} (2008) 1455
  [arXiv:0709.4635 [hep-ph]].
  
  
\bibitem{Son:1999cm}
  D.~T.~Son and M.~A.~Stephanov,
  Phys.\ Rev.\ D {\bf 61} (2000) 074012
  [hep-ph/9910491];
{\it ibid.} 
  {\bf 62} (2000) 059902
  [hep-ph/0004095];
  M.~Rho, A.~Wirzba and I.~Zahed,
  Phys.\ Lett.\ B {\bf 473} (2000) 126
  [hep-ph/9910550];
  S.~R.~Beane, P.~F.~Bedaque and M.~J.~Savage,
  Phys.\ Lett.\ B {\bf 483} (2000) 131
  [hep-ph/0002209];
  N.~Yamamoto, M.~Tachibana, T.~Hatsuda and G.~Baym,
  Phys.\ Rev.\ D {\bf 76} (2007) 074001
  [arXiv:0704.2654 [hep-ph]].
  
\bibitem{Kleinhaus:2007ve}
  V.~Kleinhaus, M.~Buballa, D.~Nickel and M.~Oertel,
  Phys.\ Rev.\ D {\bf 76} (2007) 074024
  [arXiv:0707.0632 [hep-ph]].

\bibitem{Ding:2015ona}
A review including the comparison between the hadron resonance gas and the lattice results,
  H.~T.~Ding, F.~Karsch and S.~Mukherjee,
  Int.\ J.\ Mod.\ Phys.\ E {\bf 24} (2015) no.10,  1530007
  [arXiv:1504.05274 [hep-lat]].

\bibitem{Kojo:2016dhh}
  T.~Kojo,
  Phys.\ Lett.\ B {\bf 769} (2017) 14
  [arXiv:1610.05486 [hep-ph]].
  
\bibitem{Beth:1937zz}
  E.~Beth and G.~Uhlenbeck,
  Physica {\bf 4} (1937) 915.
  
\bibitem{Dashen:1969ep}
  R.~Dashen, S.~K.~Ma and H.~J.~Bernstein,
  Phys.\ Rev.\  {\bf 187} (1969) 345.

\bibitem{Zhuang:1994dw}
  P.~Zhuang, J.~Hufner and S.~P.~Klevansky,
  Nucl.\ Phys.\ A {\bf 576} (1994) 525;
  H.~Abuki,
  Nucl.\ Phys.\ A {\bf 791} (2007) 117
  [hep-ph/0605081];
  K.~Yamazaki and T.~Matsui,
  Nucl.\ Phys.\ A {\bf 913} (2013) 19
  [arXiv:1212.6165 [hep-ph]];
  D.~Blaschke, M.~Buballa, A.~Dubinin, G.~Roepke and D.~Zablocki,
  Annals Phys.\  {\bf 348} (2014) 228
  [arXiv:1305.3907 [hep-ph]].
  
  \bibitem{nicer}  K. C. Gendreau et al.,  
Proc. SPIE 9905, Space Telescopes and Instrumentation 2016: Ultraviolet to Gamma Ray, 99051H (July 22, 2016).


\bibitem{GEO} J. Hough et al. 
available at eprints.gla.ac.uk/114852/7/114852.pdf.

\bibitem{Kagra}  Y. Aso et al. 
Phys Rev D {\bf88}, 043007:1-15 (2013).

\end{thebibliography}
\end{document}